# Bi-Directional Multi-Antenna Relay Communications with Wireless Network Coding

Chau Yuen, Woon Hau Chin
Institute for Infocomm Research (I²R), Singapore
{cyuen, whchin}@i2r.a-star.edu.sg

Yong Liang Guan, Wenhua Chen, Taoyi Tee
Nanyang Technological University, Singapore
eylguan@ntu.edu.sg

**Abstract:** In this paper, we consider a two-way or bi-directional communications system with a relay equipped with multiple antennas. We show that when the downlink channel state information is not known at the relay, the benefit of having additional antennas at the relay can only be obtained by using decode and forward (DF) but not amplify and forward (AF). The gain becomes significant when we employ transmit diversity together with wireless network coding. We also demonstrate how the performance of such system can be improved by performing antenna selection at the relay. Our results show that if downlink channel state information is known at the relay, network coding may not provide additional gain than simple antenna selection scheme.

## I. INTRODUCTION

Relay networks was first studied in the seminal work [1], where relays were studied from an information theoretic point of view. In recent years, the relay networks have become more important in communication systems as it is able to extend the communication range and improve the performance of wireless systems. In recent studies, it was shown that wireless network coding [2][3] can greatly improve the throughput of relay systems if the communication traffic is two-way or bi-directional. In this paper, we consider a relay communications system that is equipped with multiple antennas at the relay.

By having additional antenna at the relay, the problem becomes interesting as there are more options available at the relay. For example, the relay can perform network coding at the bit level with the decode and forward (DF) scheme, rather than perform network coding at the symbol level with the amplify and forward (AF) scheme.

In addition, by having multiple antennas at the relay node, various MIMO transmission schemes can be used to achieve a better decoding performance. In this paper, we consider the use of Alamouti open-loop space-time block code (STBC) [4], as well as the case when some form of channel state information (CSI) is available at the relay. We also investigate use of antenna selection when we forward the message at the relay.

We consider a communications system with a single relay equipped with two antennas, while the source and destination are each equipped with a single antenna. Simulation results show that when no CSI is available at the relay, the scheme that forward bit-level network coded bits in Alamouti STBC is the best. However, when CSI is available at the relay, simple antenna selection is good enough, and network coding does not provide any advantage.

The organization of this paper is as follows: we discuss the signal and system model in Section II, followed by the transmission schemes considered for the multi-antenna relay in Section III. We show the simulation results at Section IV and conclude the paper in Section V.

## II. SIGNAL MODEL

Consider a three-node network as shown in Figure 1, where nodes A and C, each with one antenna, would like to exchange messages. The message from A to C is denoted as $x_A$, while the message from C to A is denoted as $x_C$. The nodes A and C are assumed to be outside the communication range of each other. A relay node, B, equipped with two antennas is in between them, such that node B is able to listen to the transmission from A and C simultaneously, processes the signal according to the schemes that will be described in Section III, and then forwards the signal to nodes A and C. In this case, we denote $\mathbf{h}_{AB}$ and $\mathbf{h}_{CB}$ as the channels between nodes A and B and nodes C and B respectively. It is assumed that they operate in the same frequency band, while the uplink and downlink channel are reciprocal to which another for simplicity. However, we do not imply the nodes A, B or C know the CSI unless stated explicitly.

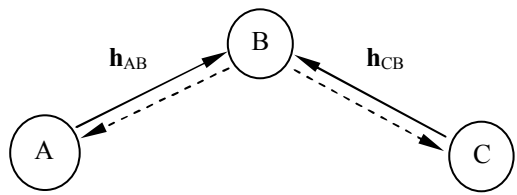

Figure 1    A three-node relay network

We assume that the communication system operates in two stages. In the first stage, the nodes A and C transmit while the relay node B is in receive mode. In the second stage, relay node B transmits while nodes A and C receive. In the first stage, the received signal at the relay node B, $\mathbf{y}_B$, can be modeled as

$$y_B = h_{AB}x_A + h_{CB}x_C + n_B$$

$$= \begin{bmatrix} h_{AB} \\ h_{CB} \end{bmatrix} \begin{bmatrix} x_A \\ x_C \end{bmatrix} + n_B \qquad (1)$$

$$= H \begin{bmatrix} x_A \\ x_C \end{bmatrix} + n_B$$

where $x_A$ and $x_C$ are the signals transmitted from nodes A and C respectively.

Depending on the scheme selected, relay node B can choose to either amplify and forward $y_B$, or estimate the signals $x_A$ and $x_C$ using the MMSE estimator as follows:

$$\begin{bmatrix} \hat{x}_A \\ \hat{x}_C \end{bmatrix} = \left( H^* H + \alpha I \right)^{-1} H^* y_B$$

$$= \left( \begin{bmatrix} h_{AB} \\ h_{CB} \end{bmatrix}^H \begin{bmatrix} h_{AB} \\ h_{CB} \end{bmatrix} + \sigma_B^2 I \right)^{-1} \begin{bmatrix} h_{AB} \\ h_{CB} \end{bmatrix}^H y_B \qquad (2)$$

where $\hat{x}_A$ and $\hat{x}_C$ are the estimated symbols for $x_A$ and $x_C$.

## III. MULTI-ANTENNA RELAYING

In this paper, we consider two scenarios, the first case is where no downlink CSI is available at the relay, and the second case is where there is some form of downlink CSI available at the relay, *i.e.* the multi-antenna relay knows which of its downlink channel has a larger gain, hence antenna selection can be performed.

### A. No CSI at the relay

#### *Amplify and Forward (AF)*

The relay node B can use various techniques to forward the signal received to nodes A and C. If no CSI is available at the relay, the simplest way of amplify and forward (AF) scheme is for node B to forward

$$s_B = \frac{y_B}{\sqrt{\left| h_{AB} \right|^2 + \left| h_{CB} \right|^2 + 2\sigma_B^2}} = \alpha_B y_B \qquad (3)$$

where $\alpha_B$ is the power normalization constant, and $\sigma_B^2$ is the noise power at the each of the antennas of node B. Since node B has two antennas, hence, it has twice the noise power as compared to the case of a single-antenna relay. It will be shown later on, that by using AF, the relay with two-antenna suffers a slight degradation as compared to single-antenna case.

At node A, $x_C$ can be estimated as follows since $x_A$ is known to A. Such a scheme is commonly known as "*analog network coding*":

$$y_A = h_{BA}s_B + n_A$$

$$= \alpha_B h_{BA} \left( h_{AB}x_A + h_{CB}x_C + n_B \right) + n_A \qquad (4)$$

$$\hat{x}_C = \left( \frac{y_A}{\alpha_B} - h_{BA}h_{AB}x_A \right) \Big/ \left( h_{BA}h_{CB} \right)$$

It is assumed that the node has the necessary channel knowledge to perform the decoding, as the objective of this study is to investigate performance of various schemes, and leave the practical aspect of channel estimation for future works.

#### *Decode and Forward with Spatial Multiplexing (DF-SM)*

Besides AF, since node B has two antennas, it is able to decode the signals $x_A$ and $x_C$ before it forward them to nodes C and A respectively. In this paper, we assume nodes B estimates the signals using the MMSE estimator as in (2).

The estimated signals are then decoded and re-encoded, and sent out as the spatially multiplexed signal

$$s_B = \begin{bmatrix} \lfloor \hat{x}_A \rfloor \\ \lfloor \hat{x}_C \rfloor \end{bmatrix} \qquad (5)$$

where $\lfloor \ \rfloor$ represents the decoding and re-encoding operation on the estimated symbols. The estimation of the intended signals at node A is as follows:

$$y_A = h_{BA}s_B + n_A$$

$$= h_{BA} \begin{bmatrix} \lfloor \hat{x}_A \rfloor \\ \lfloor \hat{x}_C \rfloor \end{bmatrix} + n_A \qquad (6)$$

$$\hat{x}_C = \left( y_A - h_{BA,1}x_A \right) \Big/ h_{BA,2}$$

where $h_{BA,i}$ is the *i*th element of the channel vector. Since node A does not know if there is any estimation error at the relay node, the cancellation may not be perfect. Likewise node C can estimate $x_A$ accordingly.

#### *Decode and Forward with Network Coding (DF-NC)*

We can also perform network coding at bit level at the relay node B by first demodulating the detected signals into two bits steams, $b_A$ and $b_C$:

$$\hat{b}_A = \text{demod}\left( \hat{x}_A \right) \qquad \hat{b}_C = \text{demod}\left( \hat{x}_C \right)$$

$$x_B = \text{mod}\left( \hat{b}_A \oplus \hat{b}_C \right) \qquad (7)$$

where $\oplus$ is the XOR process, mod() and demod() are the modulation and demodulation functions, and $x_B$ is the new stream of data produced by XORing the estimated bit streams $b_A$ and $b_C$. Relay node B can then forward the new streams of data as

$$s_B(t) = \begin{bmatrix} x_B(t) \\ x_B(t) \end{bmatrix}. \qquad (8)$$

The estimation of the intended signals at nodes A is as follows:

$$y_A = h_{BA}s_B + n_A$$

$$= h_{BA} \begin{bmatrix} 1 \\ 1 \end{bmatrix} x_B(t) + n_A \qquad (9)$$

$$\hat{x}_B = y_A \Big/ \left( h_{BA} \begin{bmatrix} 1 \\ 1 \end{bmatrix} \right)$$

where $\hat{x}_B$ is the estimation of the network coded symbol transmitted from the relay. The intended message from node C can then be recovered by performing XOR operation on the network coded bit stream, $\hat{b}_B$, with the known message $b_A$:

$$\hat{b}_C = \hat{b}_B \oplus b_A. \qquad (10)$$

*Decode and Forward with Network Coding and Alamouti Coding (DF-NC-Alamouti)*

To exploit the space time resources of the multiple antennas at the relay, we can also encode (7) in using the Alamouti space time block code [3]. As a result, the transmitted data signal is

$$\mathbf{s}_B(t) = \begin{bmatrix} x_B(t) & -x_B^*(t+1) \\ x_B(t+1) & x_B^*(t) \end{bmatrix}. \tag{11}$$

The decoding of the Alamouti STBC can be referred to [4]. The estimated bit stream can be recovered as in (10), hence will not be discussed in detail.

## B. With CSI

In this part, we consider that there is enough information at the relay such that antenna selection can be performed.

*Decode and Forward with Antenna Selection (DF-ANT)*

After the relay estimated $x_A$ and $x_C$ from (2), the relay will forward $\hat{x}_C$ only through the stronger channel of $\mathbf{h}_{BA}$, likewise forward $\hat{x}_A$ only through the stronger channel of $\mathbf{h}_{BC}$. The details of the forwarding scheme are summarized in Table 1.

Table 1  The forwarding scheme for DF-ANT

| If  $\mathbf{h}_{BA}(1) > \mathbf{h}_{BA}(2)$  &  $\mathbf{h}_{BC}(1) > \mathbf{h}_{BC}(2)$ | $\begin{bmatrix} \lfloor \hat{x}_A \rfloor + \lfloor \hat{x}_C \rfloor \\ 0 \end{bmatrix}$ |
|---|---|
| If  $\mathbf{h}_{BA}(2) > \mathbf{h}_{BA}(1)$  &  $\mathbf{h}_{BC}(2) > \mathbf{h}_{BC}(1)$ | $\begin{bmatrix} 0 \\ \lfloor \hat{x}_A \rfloor + \lfloor \hat{x}_C \rfloor \end{bmatrix}$ |
| If  $\mathbf{h}_{BA}(1) > \mathbf{h}_{BA}(2)$  &  $\mathbf{h}_{BC}(2) > \mathbf{h}_{BC}(1)$ | $\begin{bmatrix} \lfloor \hat{x}_A \rfloor \\ \lfloor \hat{x}_C \rfloor \end{bmatrix}$ |
| If  $\mathbf{h}_{BA}(2) > \mathbf{h}_{BA}(1)$  &  $\mathbf{h}_{BC}(1) > \mathbf{h}_{BC}(2)$ | $\begin{bmatrix} \lfloor \hat{x}_C \rfloor \\ \lfloor \hat{x}_A \rfloor \end{bmatrix}$ |

*Decode and Forward with Network Coding and Antenna Selection (DF-NC-ANT)*

After the relay encoded the estimated bit streams as $x_B$ as shown in (7), $x_B$ will transmit using only the first antenna of the relay if the first antenna has a better channel gain than the second antenna to both nodes A and C; likewise $x_B$ will transmitted only at the second antenna of the relay if the second antenna has a better channel gain than the first antenna to both nodes A and C. If none of the antenna has a clear gain over another, then $x_B$ will be transmitted on both antennas as similar to (8).

Table 2  The forwarding scheme for DF-NC-ANT

| If  $\mathbf{h}_{BA}(1) > \mathbf{h}_{BA}(2)$  &  $\mathbf{h}_{BC}(1) > \mathbf{h}_{BC}(2)$ | $\begin{bmatrix} x_B \\ 0 \end{bmatrix}$ |
|---|---|
| If  $\mathbf{h}_{BA}(2) > \mathbf{h}_{BA}(1)$  &  $\mathbf{h}_{BC}(2) > \mathbf{h}_{BC}(1)$ | $\begin{bmatrix} 0 \\ x_B \end{bmatrix}$ |
| If  $\mathbf{h}_{BA}(1) > \mathbf{h}_{BA}(2)$  &  $\mathbf{h}_{BC}(2) > \mathbf{h}_{BC}(1)$ | $\frac{1}{\sqrt{2}} \begin{bmatrix} x_B \\ x_B \end{bmatrix}$ |
| If  $\mathbf{h}_{BA}(2) > \mathbf{h}_{BA}(1)$  &  $\mathbf{h}_{BC}(1) > \mathbf{h}_{BC}(2)$ | $\frac{1}{\sqrt{2}} \begin{bmatrix} x_B \\ x_B \end{bmatrix}$ |

## IV.  SIMULATION RESULTS

We consider a Rayleigh flat fading relay network with two-way or bi-directional communications. We assume all the nodes are synchronous, and they all have the same SNR with all the nodes employing QPSK modulation.

We compare the BER performance in Figure 2, where the blue dotted lines represent the results of AF, the black solid curves represent the DF scheme without CSI information at the relay, while the red triangular and red square lines are the DF scheme with antenna selection.

It can be seen from Figure 2 that with AF, relay with two antennas provide no gain over relay with one antenna. The performance is in fact slightly worse than one antenna case, this could be due to the power imbalance of the transmitted signal. Additionally, during transmission, we restrict the total transmission power to be same as the case with one antenna, hence, part of the power is being used to forward the noise, and this leads to the slight degradation in the performance.

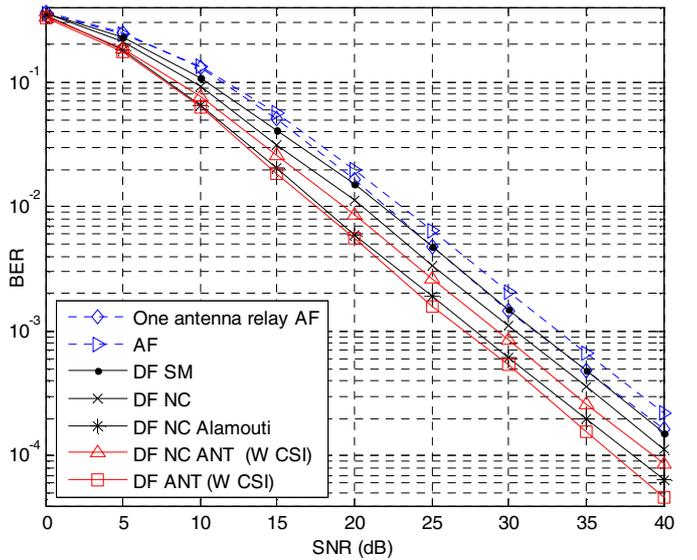

Figure 2  Simulated BER for various schemes

Next, with DF, the performance is significantly improved, and the gain provided by the additional antenna becomes visible. We show that DF-NC is better than DF-SM, and DF-NC-Alamouti is better than DF-NC. This is because by using network coding, the gain becomes even more significant as the two data streams are combined into one single stream using XOR. Therefore all the transmission power can be concentrated on this single stream, rather than shared between two streams. And it can be seen that with the use of the Alamouti transmit diversity scheme, it provides the best BER performance among all scheme without CSI.

It should be noted that, for all the schemes, including the Alamouti scheme, they have the same diversity gain (as in the slope of the curve). This suggests that the diversity order of such system is being capped by uplink.

However, for antenna selection, we get a different conclusion from previous case where no CSI is available at relay. It can be seen that DF-ANT performs better than DF-NC-ANT. In other words, network coding may not provide any additional gain when antenna selection is performed. This can be explained as follows: when the CSI is known at the relay, it is best to simply concentrate the transmission power on the desired direction. For the case of network coding, two symbols are "coded" together, hence it would be a kind of waste of energy when the channel of both side do not match to the same antenna.

## V. CONCLUSION

We compared several schemes for a two-way / bi-directional relay communications system that has two antennas at the relay node. We show that by using simple amplify and forward, the multiple antennas at the relay provide no additional benefit. On the other hand, by combining network coding with decode and forward, the multiple antennas can provide additional gain, and the gain becomes significant when Alamouti open-loop transmit diversity is employed. When CSI is available at the relay, such as when antenna selection is performed, we show that network coding may not be an optimal solution. As for the future work, depending on the amount and quality of the channel state information available at the source and relay, more complicated schemes can be developed.